\documentstyle[12pt]{article}

\catcode`\@=11

\def\eqbegin         {  \begin{eqnarray}  }
\def\eqend           {  \end{eqnarray}  }
\def\beq{\begin{equation}}
\def\eeq{\end{equation}}
\def\sectionnumbering 
{\setcounter{equation}{0}\renewcommand{\theequation}
{\arabic{section}.\arabic{equation}}}
\def\lamda        { \lambda }
\def\iPhi         { {\mit \Phi}}  
  
\def\iDelta       { {\mit \Delta}}
\def\iLamda       { {\mit \Lambda}}

\def\I{{\bf Im}\ }
\def\R{{\bf Re}\ }
\def\N{{\bf N}\ } 
\def\Z{{\bf Z}}
\def\CVO#1#2#3{\!\left( \matrix{ #1 \cr #2 \ #3 \cr} \right)\!}

\def\barr{\begin{eqnarray}}
\def\earr{\end{eqnarray}}

\def\u1{\widehat{U(1)}}
\def\su2{\widehat{SU(2)}_1}

\def\so2n{$SO(2n)$}

\def\rep{representation }

\def\hsf{\hspace{4mm}}

\title{Modular Invariants in the Fractional Quantum Hall Effect}
\author{Kazusumi Ino\thanks{e-mail:ino@kodama.issp.u-tokyo.ac.jp}}
\date{}

\begin{document}
\markright{}
\thispagestyle{empty}
\thispagestyle{empty}
\begin{titlepage}
\begin{center}
\hfill  \quad Aug-98 \\
\vskip .5 in
{\LARGE Modular Invariants}
\vskip 0.1in
{\LARGE in} 
\vskip 0.1in
{\LARGE  the Fractional Quantum Hall Effect }
\vskip 0.2in
KAZUSUMI INO \\
{\em Institute for Solid State Physics, University of Tokyo,} \\
{\em Roppongi
 7-22-1,  Minatoku,  Tokyo,  106, Japan}  \\ 
{\sf ino@kodama.issp.u-tokyo.ac.jp}
\end{center}
\vskip .1 in

\begin{abstract}
We  investigate the modular properties of the characters 
which appear in the partition functions of nonabelian 
fractional quantum Hall states. 
We first give the annulus partition function for 
nonabelian FQH states formed by spinon and holon (spinon-holon state).
The degrees of freedom of spin are described by  
the affine $SU(2)$ Kac-Moody algebra at level $k$.  The partition function 
and the Hilbert space of the edge excitations 
decomposed differently according to whether $k$ is even or odd. 
We then investigate the full modular properties of 
the extended characters for nonabelian fractional quantum Hall states. 
We  explicitly  verify the 
modular invariance of the annulus grand partition functions for 
spinon-holon states, the Pfaffian state and the 331 states.  
This enables one to extend 
the relation between the modular behavior and the topological order 
to nonabelian cases. 
For the Haldane-Rezayi state, we find that 
 the extended characters do not form a representation of
 the modular group, thus the modular invariance is broken.    
We also find a new relation between the Haldane-Rezayi state and 
the 331 state and suggest its implications for 
'The $\nu=\frac{5}{2}$ Enigma'. 
\end{abstract}

\pagenumbering{arabic}

\end{titlepage}
\section{Introduction}
Fractional quantum Hall (FQH) states \cite{review,laugh} have edge excitations 
as their low energy excitations \cite{stone,wen}.      
Some experiments have already  observed characteristic behaviors  
 of theoretical predictions \cite{kane,milli}.    
The edge excitations form representations of chiral algebra 
extended by the holomorphic field for the electron. 
One way to give a complete description of the Hilbert space of 
edge excitations is to construct  the partition function.  
The partition function for 
edge excitations  has been discussed for a  disk and 
an  annulus geometry \cite{wen,cappelli,wenwu2,milo,cappelli2}.  
Especially, Cappelli and Zemba \cite{cappelli2} gave   
extended  modular invariance conditions  which should be satisfied  by 
the annulus partition function  for edge excitations of abelian  
quantum Hall states.  The annulus partition function encodes not only  
the spectrum of edge excitations but also  
the  {\it topological order} \cite{topord} 
in its modular properties through the Verlinde formula.   
They also found non-diagonal modular invariants partition functions  for 
abelian cases from the charcters of the affine $U(1)^{m}$ algebra, 
which may explain the experimentally observed plateaus beyond 
the Jain's series \cite{jain} such as $\nu=\frac{4}{11}, \frac{4}{13}$. 
Recently all the possible  $U(1)^{m}$ modular invariants for 
abelian quantum Hall effect  
were classified by Gannon \cite{gannon}.  
Also the annulus partition functions for the paired states 
(Pfaffian \cite{moore}, Haldane-Rezayi \cite{halrez})  
were derived in \cite{milo},  but the modular 
properties of these partition functions have not been worked out.

In this paper, 
we  first derive the partition functions for the  
edge excitations of the spin-singlet FQH state ({\bf spinon-holon} state) 
constructed from conformal blocks of the $SU(2)$ Wess-Zumino-Witten 
model at level $k$.  
This state is an example of nonabelian FQH states \cite{moore} when $k>1$. 
The statistical property of the quasiparticle in this state 
was discussed by Blok and Wen \cite{blok}.  
The case of $k=1$ arises from the Jain's hierarchy 
\cite{spin,martino,cappelli2,cabra}.  
It also gives  the spin-singlet Halperin state \cite{halp,bala1}.    
In these models, electrons or composite fermions 
 form a spin $k/2$ representation 
of $SU(2)$.  The  degeneracy  may arise from 
spin,  multi-layer, degenerate bands of composite fermions and so on.  
We examine the edge excitations of spinon-holon states 
from the bulk wave functions 
obtained from the bulk CFT \cite{moore,cris}, which is 
 based on the general relation  between the bulk CFT and Chern-Simons 
theory \cite{witten}.  
  According to that, 
edge states can be treated like states on punctures in the 
thermodyamic limit, which is also supported by a 
numerical study \cite{wenwu2}.
We find  that the partition function or 
the Hilbert space of the edge excitations is decomposed 
differently according 
to whether $k$ is even or odd.  Accordingly, 
the bulk excitation of the state have 
the different quasiparticle spectrum in the holon's part. 

We then investigate the modular properties of the characters which 
appear in annulus partition functions for nonabelian FQH states
including the spinon-holon states and the paired states 
(Pfaffian, Haldane-Rezayi, 331).  
These states are of great interest since they may be realized in 
the even-denominator plateaus in a single and double-layers systems 
\cite{willet,he,morf}.  Among them, the  $\nu=\frac{5}{2}$ plateau 
 \cite{willet} remains an 'enigma' \cite{eisen}.   Experiments support that 
the state is the spin-singlet Haldane-Rezayi state. 
Numerical studies, however,  show that  
the spin-polarized Pfaffian-like state is more favored   
\cite{morf}.     
Recently nonabelian statistics 
in these paired states have attracted considerable attentions 
\cite{nayak,gurarie,fradkin,schoutens}.
Although the modular properties of the Haldane-Rezayi state are 
discussed in Ref.\cite{gurarie},   the full modular properties 
including the charge sectors of nonabelian FQH states 
have not been addressed. 
We derive the full modular  matrix elements
 of extended characters and explicitly verify  
the extended  modular invariance conditions 
 for spinon-holon states, the Pfaffian states and the 331 states.  
For the Haldane-Rezayi state, we find
 that the extended characters do not form a representation 
of the modular group.   
This implies  that 
the annulus partition functions cannot be modular invariant and
requires some modification.  We give a new  way to 
cure this pathologic behavior by establishing new relation 
between the Haldane-Rezayi state and the 331 state.
 
For the other nonabelian FQH states, 
the modular invariance 
 enables one to extend the relation between the topological order 
and modular properties  of Cappelli and Zemba.

The organization of the paper is as follows. 
In Section 2, we derive the partition functions for 
 the spinon-holon FQH state using the method of Ref.\cite{ino}. 
In Sec. 3, we determine the modular properties of 
the characters of nonabelian FQH states, and 
investigate the extended modular invariance conditions for them. 
 Sec. 4  is devoted to conclusions.

\section{Spinon-Holon FQH States} 
\sectionnumbering
We consider the FQH states 
formed by {\bf spinon} and {\bf holon}.  
The degrees of freedom of spin is carried by spinon, given by 
the affine $SU(2)_k$  algebra.   The degrees of freedom of holon 
is carried by holon, given by the affine $U(1)$ Kac-Moody algebra.

Let us first recall the representation of $SU(2)_k$.  
$SU(2)_k$ has $k+1$  integrable representations 
$[\phi_l]$  with $l=0,\cdots,k$, 
namely the ones with $SU(2)$ isospin $\frac{1}{2}l\le
\frac{1}{2}k$. The conformal weight for $\phi_l$ is 
\eqbegin 
h_l=\frac{l(l+2)}{4(k+2)}.
\eqend 
 The fusion rules are \cite{gepner}
\eqbegin 
\phi_l \times \phi_{l'} = 
\sum_{j=|l- l'|}^{{\rm min}(l+l',2k-l-l')} \phi_j ,
\label{su2fusion}
\eqend   
where $j-|l-l'|$ is an even integer.  
This model is supposed to have 
a quasiparticle with nonabelian statistics when $k>1$ \cite{blok}.  

\paragraph{Disk}
We first consider the edge excitations for a disk geometry. 
We can form the ground-state wave function from the primary fields in 
$SU(2)_k$ by coupling a chiral boson $\varphi$.    
Since the electrons must have fermionic statistics, 
they are constructed from a so-called simple current i.e. whose
operator product expansion with itself only gives rise to  one
representation.   
From (\ref{su2fusion}),  the only simple current in this model is the 
representation $[\phi_k]$. 
 Since $\phi_k$ has a conformal weight 
$h_k=\frac{k}{4}$,  one couples the field $ e^{i\sqrt{q}\varphi}$ with 
$q=\frac{k}{2}+s$ to form a well defined wave function.  
Here $s+k$ must be an odd integer to account for the 
fermionic statistics of electron. This can be seen by using 
fermionic construction of $SU(2)_k$ WZW model \cite{blok}.  
The primary field for the electron is then, 
\eqbegin 
V^{\alpha}_{k}e^{i\sqrt{q}\varphi}, 
\label{fel}
\eqend 
where $V^{\alpha}_{k}, \hspace{2mm} 
\alpha=-k,-k+1, \cdots,k $ are the vertex operators of $[\phi_k]$.
We can also take $e^{-i\sqrt{q}\varphi}$ in (\ref{fel}). In that case, 
we refer to the number of electrons as $-N$ and $\varphi$ 
would be  replaced by $-\varphi$ in the following. 
From these operators, we can construct the  ground-state wave function for 
a fractional quantum Hall state of $N$ $SU(2)$ electrons on a disk: 
\eqbegin 
\langle \Psi_{\rm edge}^{\vee}| 
\prod_{i=1}^{N}V^{\alpha_i}_{k}e^{i\sqrt{q}\varphi}|vac \rangle 
{\rm exp}
\left[\sum_i -\frac{1}{4}|z_i|^2\right],  
\label{su2k}
\eqend 
where $\alpha_1,\cdots \alpha_N $ are indices for  
the spin of spinons and $\Psi_{\rm edge}$ is given by 
\eqbegin 
 \Psi_{\rm edge} &=& e^{-iN\sqrt{q}\varphi},  \hsf 
h_{\rm edge}=\frac{1}{2}N^2q.  \hspace{4mm}  {\rm for} \hspace{2mm} N \hspace{3mm} {\rm even},  \\
\Psi_{\rm edge} &=& V^{\alpha_{\infty}}_{k} 
e^{-iN\sqrt{q}\varphi}, \hsf  
h_{\rm edge}=\frac{k}{4}+\frac{1}{2}N^2q, \hspace{4mm}   {\rm for} \hspace{2mm} N
\hspace{3mm} {\rm    odd}, \\
 \langle \Psi_{\rm edge}^{\vee}| &=&  \lim_{z_{\infty}\rightarrow
   \infty} 
\langle vac|
 \Psi_{\rm edge}(z_{\infty}) z_{\infty}^{2h_{\rm edge}}. 
\label{Boundary} 
\eqend             
This state has a filling fraction $\nu=\frac{1}{q}$.

Let us  especially consider  $SU(2)_1$.  
 $SU(2)_1$ WZW model has two primary fields, 
\eqbegin 
V^{\downarrow}(z_{\downarrow}), \hspace{4mm} 
V^{\uparrow}(z_{\uparrow}).
\eqend 
The conformal weights for these vertex operators are   
$\frac{1}{4}$.
We can construct a $SU(2)$ singlet ground state from these fields 
($N$:even): 
\eqbegin 
\langle \prod_{i=1}^{N}V^{\uparrow}({z_i}^{\uparrow})
e^{i\sqrt{n+\frac{1}{2}}\varphi({z_i}^{\uparrow})}
\prod_{i=1}^{N}V^{\downarrow}(z_i^{\downarrow})
e^{i\sqrt{n+\frac{1}{2}}\varphi(z_i^{\downarrow})}\rangle .
\label{halperin}
\eqend 
This wave function is manifestly $SU(2)$ invariant. It is calculated to be 
\eqbegin 
\Phi_{\rm Halp}= 
\prod_{i<j}(z_i^{\uparrow}-z_j^{\uparrow})^{n+1}\prod_{i<j}(z_i^{\downarrow}-z_j^{\downarrow})^{n+1}\prod_{i<j}(z_i^{\uparrow}-z_j^{\downarrow})^{n}{\rm exp}\left
[ -\frac{1}{4}\sum(|z^{\uparrow}_i|^{2}+|z^{\downarrow}_i|^{2})\right]. 
\nonumber\\
\eqend 
This is  the spin-singlet state 
called the Halperin state \cite{halp}. In this  case, 
the spinon is {\it semion}.     

For general $k$, the spectrum of 
excitations is restricted by the requirement of 
single-valuedness and non-singularity of wave function 
in electron coordinates. Since 
$V^{\alpha_1}_{j} \hspace{2mm} (j=1,\cdots,k)$ and $V^{\alpha_2}_k$ have an operator product expansion  
\eqbegin 
V^{\alpha_1}_{j}(z_1)V^{\alpha_2}_k(z_2) \sim 
\frac{C_{\alpha_1\alpha_2}^{\alpha}}{(z_1-z_2)^{\frac{j}{2}}}V^{\alpha}_{k-j}(z_2)+\cdots,  
\eqend
where $C_{\alpha_1\alpha_2}^{\alpha}$ is a constant, 
the allowed couplings are 
\eqbegin 
V_{j}e^{i\frac{(2n+j)}{2\sqrt{q}}\varphi}, \hspace{4mm} n \in \N, 
\hspace{3mm}{\rm for}\ \ j {\rm : odd}, 
\label{su2coup1}\\ 
V_{j}e^{i\frac{n+j/2}{\sqrt{q}}\varphi},  n\in \N, \hspace{3mm} 
{\rm for}\ \ j {\rm : even}.
\label{su2coup2}  
\eqend 
To see the relation between the bulk and the edge explicitly, 
 let us use the chiral vertex operator (CVO) 
expression of FQH state \cite{ino}. 
We  denote the CVO :
$[\phi_{j_3}] \rightarrow {\rm Hom}([\phi_{j_1}] \rightarrow 
[\phi_{j_2}])$ 
for $SU(2)_k$ WZW model as
\eqbegin 
\iPhi^{\alpha}\CVO{j_3}{j_2}{j_1}(z) .
\eqend 
For our purpose,  the CVO  represents the annulus with 
the representation  $[\phi_{j_1}]$ 
inserted at $z$ with the inner and outer edge states 
$[\phi_{j_2}]$ and   $[{\phi_{j_3}}^{\vee}]$ respectively \cite{ino}.

Suppose that we take  a disk-like region $D_1$ in the  disk,
 which contains only one  electron. 
Then we make a series of annulus-like regions $D_2\cdots D_N$,  
each of which contains  one electron respectively such 
that the outer region of $D_N$ is the boundary of the disk. 
One can correspond  $D_1, \cdots D_N$ to CVOs.  
Then the ground-state wave function (\ref{su2k}) can be 
 expressed as,  for even $N$: 
\eqbegin 
\widetilde{\iPhi}_{\rm disk}=
 \iPhi^{\alpha_1}\CVO{k}{0}{k}(z_1)
 \iPhi^{\alpha_2}\CVO{k}{k}{0}(z_2)
 \cdots 
 \iPhi^{\alpha_{N-1}}\CVO{k}{0}{k}(z_{N-1})
 \iPhi^{\alpha_N}\CVO{k}{k}{0}(z_N).
\eqend 
for odd $N$: 
\eqbegin 
\widetilde{\iPhi}_{\rm disk} =\iPhi^{\alpha_1}\CVO{k}{k}{0}(z_1)
 \iPhi^{\alpha_2}\CVO{k}{0}{k}(z_2)
 \cdots 
 \iPhi^{\alpha_{N-1}}\CVO{k}{0}{k}(z_{N-1})
 \iPhi^{\alpha_N}\CVO{k}{k}{0}(z_N),  
\eqend
Here we have omitted the indices for the rational torus.

The edge excitations are generated on the field at infinity,   
 $\Psi_{\rm edge}$. It is $V^{\alpha_{\infty}}_0$ for $N$ even and 
$V^{\alpha_{\infty}}_k$ for $N$ odd.  
Edge excitations are generated by current algebras as   
\eqbegin 
J^{a_1}_{-n_1} \cdots J^{a_{l_1}}_{-n_{l_1}} 
\widetilde{j}_{-n_{1}'}\cdots \widetilde{j}_{-n_{l_2}'}\Psi_{\rm edge}, 
\eqend
where $\widetilde{j_{n}}$ is the generator of the 
affine $U(1)$ Kac-Moody algebra 
\eqbegin
 [ \widetilde{j}_n,\widetilde{j}_m ]&=& n \delta_{m+n}, 
\label{U(1)KacMoody}
\eqend
and $J^{a}_{n}$ are  the generators of the 
 affine $SU(2)_k$  Kac-Moody algebra satisfying 
\eqbegin 
[J^{a}_{m},J^{b}_{n}]=i\epsilon^{abc}J^{c}_{m+n}+km\delta^{ab}\delta_{m+n,0}. 
\label{KacMoody}
\eqend 
Here $\epsilon^{abc}$ is a rank 3  antisymmetric tensor.
We denote CVO expressions for the edge excited state as 
$\widetilde{\iPhi}^{(n_1,n_2,\cdots)}$.    

We define an operator $M$ on chiral vertex operators by 
\eqbegin 
M\iPhi^{\alpha}\CVO{j_3}{j_2}{j_1}=(-\iDelta_{j_1}+\iDelta_{j_3}-\iDelta_{j_2})\iPhi^{\alpha}\CVO{j_3}{j_2}{j_1}(z),   
\eqend 
where $\iDelta_{j_m}$ is the conformal weight of the field 
$\phi_{j_m}$.  The action on the expressions with descendant field 
is defined similarly. 
 We demand that it satisfies the Leibnitz rule when it acts
 on the product of chiral vertex operators.   
As shown in \cite{ino}, 
$M$  acts as the angular-momentum operator on these states  
when it acts on the monomials  
$\widetilde{\iPhi}^{(n_1,n_2,\cdots)}$.   
Let us denote the vector space 
spanned by such monomials  as $\Omega_{\rm edge}$.
Then the partition function for the state on the disk is given by 
\eqbegin 
Z^{\rm disk}(\tau)=
{\rm Tr}_{\Omega_{\rm edge}}\left({\rm e}^{ 
 2\pi  i\tau(M-\frac{c}{24})}\right)
&=&\frac{\omega^{M_0}\chi^{(k)}_0(\tau)}{\eta(\tau)}
\hspace{4mm}  {\rm for} \hspace{3mm} N \hspace{3mm} {\rm even},\\ 
M_0&=&\frac{1}{2}qN(N-1)-\frac{k}{4}N, \\
Z^{\rm disk}(\tau)&=&\frac{\omega^{M_0}\chi^{(k)}_{k}(\tau)} 
{\eta(\tau)}\hspace{4mm} {\rm for}\hspace{3mm} N \hspace{3mm} 
{\rm odd}, \\
c=\frac{3k}{k+2}+1, 
\hspace{4mm} M_0&=&\frac{1}{2}qN(N-1)-\frac{k}{4}(N-1). 
\eqend 
Here $\eta$  is the Dedekind function
\eqbegin 
\eta(\tau)=\omega^{\frac{1}{24}}\prod_{n=1}^{\infty}(1-\omega^n), 
\eqend 
where $\omega=e^{ 2\pi i \tau}$,   
and $\chi^{(k)}$ is the character for the affine $SU(2)$ Kac-Moody algebra at 
 level $k$, 
\eqbegin 
\chi^{(k)}_{\lamda}(\tau)=\frac{\omega^{(\lamda+1)^{2}/4(k+2)}}{\eta(\tau)^3}
\sum_{n \in \Z} (\lamda+1+2n(k+2))\omega^{n(\lamda+1+(k+2)n)}.
\eqend 
$\chi^{(k)}_{\lamda}$ can be decomposed into 
the spinon basis \cite{bouwk}, in which the number of quasiparticle 
is explicit.

\paragraph{Annulus}
Let us next consider a state on an annulus.  
We first take a disk in the annulus which contains all the electrons, 
then we take a three-holed sphere whose holes are the two  edges of 
the annulus and the boundary of the disk.  By using duality 
transformations on CVOs \cite{moore2},   it can be  expressed  as \cite{ino}, 
\eqbegin 
 \iLamda \CVO{\beta_2}{\beta_1}{\beta} \widetilde{\iPhi}_{\rm disk}
(z_1, \cdots, z_N).  
\eqend 
We will denote the vector space spanned by 
the expressions for edge excited states as 
$\Omega_{N}$.

We define the action of $M$ on $\iLamda$ by   
\eqbegin 
M\iLamda\CVO{j_1}{j_2}{j_3}=(\iDelta_{j_1}+\iDelta_{j_2}-\iDelta_{j_3})
\iLamda\CVO{j_1}{j_2}{j_3}, 
\eqend 
and  introduce another operator $\overline{M}$ by 
\eqbegin 
\overline{M}\iPhi\CVO{j_1}{j_2}{j_3}&=&(\iDelta_{j_1}+\iDelta_{j_2}-\iDelta_{j_3})\iPhi\CVO{j_1}{j_2}{j_3}, \\
\overline{M}\iLamda\CVO{j_1}{j_2}{j_3}&=&(\iDelta_{j_1}+\iDelta_{j_2}-\iDelta_{j_3})
\iLamda\CVO{j_1}{j_2}{j_3}.
\eqend 
The pseudoenergy \cite{milo} is then given by
$M_E=(M+\overline{M})/2$.  
We can also give  the total charge operator by  
$Q={2}(\overline{M}-M)/(2q+k)$ for the expressions with positive $N$ and  
$Q=2(M-\overline{M})/(2q+k)$ for the expressions with negative $N$.
Then $\beta_1$ and $\beta_2$ satisfy 
\eqbegin 
\beta_1 &=& \beta_2 \hspace{9mm}         {\rm for}\hspace{2mm}N\ \
{\rm : even}, \\ 
\beta_1 &=& k-\beta_2 \hspace{5mm} {\rm for}\hspace{2mm}N \ \ {\rm : odd}.
\label{su2c}
\eqend
Moreover we must apply the  selection rule for 
the charge sectors.  

\subparagraph{k:even}
First let us consider the case when $k$ is even.  
We denote the charge sectors for two edges as $m_1, m_2, \hspace{3mm} 
m_1,m_2 \in \Z$. Since $k$ is even,  
if $m_1$ is from the even (odd) sector, 
$m_2$ is from  the even (odd) sector and visa versa. 
They have a relation  
\eqbegin 
m_1=m_2+Nq.
\label{bb}
\eqend  
We can't determine $m_1$ and $m_2$ in this case. It is 
convenient to consider the grand canonical ensembles of electrons.  
In the Laughlin state, we gather all the charge 
sectors differing by integral charges into a single sector since 
 integer charges are from the electrons. 
In spinon-holon states,   
we must  gather all the charge sectors differing 
by {\it even} integral charges into a single sector, since electrons 
are coupled by the degree of freedom from $SU(2)_k$. 
Thus we introduce  the spaces 
$\Omega^{\rm even}_{\rm edge}=\bigoplus_{N:{\rm even}}\Omega_N$ 
and 
$\Omega^{\rm odd}_{\rm edge}=\bigoplus_{N:\rm odd}\Omega_N$ 
and   the following characters: 
\eqbegin 
\chi_{r/q}^{\rm even}(\tau,\zeta)=
{{\rm e}^{ -{\pi\over q}{\left(\I\zeta\right)^2\over \I\tau} }
\over \eta}\ \sum_{m \in Z_{\rm even}}
{\rm e}^{ 2\pi i\tau\frac{(mq+r)^2}{2q}+2\pi i\zeta(m+\frac{r}{q})}, 
\label{charaq}
\\
\chi_{r/q}^{\rm odd}(\tau,\zeta)= 
{{\rm e}^{ -{\pi\over q}{\left(\I\zeta\right)^2\over \I\tau} }
\over \eta}\ \sum_{m \in Z_{\rm odd}}
{\rm e}^{ 2\pi i\tau\frac{(mq+r)^2}{2q}+2\pi i\zeta(m+\frac{r}{q})},\\ 
\chi_{(r+1/2)/q}^{\rm even}(\tau,\zeta)=
{{\rm e}^{ -{\pi\over q}{\left(\I\zeta\right)^2\over \I\tau} }
\over \eta}\ \sum_{m \in Z_{\rm even}}
{\rm e}^{ 2\pi i\tau\frac{(mq+r+1/2)^2}{2q}+2\pi i\zeta(m+\frac{r+1/2}{q})}, \\
\chi_{(r+1/2)/q}^{\rm odd}(\tau,\zeta)= 
{{\rm e}^{ -{\pi\over q}{\left(\I\zeta\right)^2\over \I\tau} }
\over \eta}\ \sum_{m \in Z_{\rm odd}}
{\rm e}^{ 2\pi i\tau\frac{(mq+r+1/2)^2}{2q}+2\pi i\zeta(m+\frac{r+1/2}{q})}. 
\label{charaq4}
\eqend
The prefactor will be   necessary for the extended 
modular invariance conditions 
which will be explained in Sec.\ref{mic}. 
From these definitions, they satisfy 
\eqbegin 
\chi^{\rm even}_{r/q}=\chi^{\rm even}_{-r/q}=\chi^{\rm even}_{(2q+r)/q}
=\chi^{\rm odd}_{(q+r)/q}, \\
\chi^{\rm odd}_{r/q}=\chi^{\rm odd}_{-r/q}=\chi^{\rm odd}_{(2q+r)/q}
=\chi^{\rm even}_{(q+r)/q}.
\eqend

From  the couplings of (\ref{su2coup1}) and (\ref{su2coup2}), 
we see that   the characters for the allowed excitations of chiral algebra 
become ($a$=even, odd):
\eqbegin 
\chi_{j,r}^{a}&=&\chi_j^{(k)}\chi_{(r+1/2)/q}^{a}, \hspace{4mm} {\rm for}\hspace{2mm}j{\rm :odd }\\
\chi_{j,r}^{a}&=&\chi_{j}^{(k)}\chi_{r/q}^{a} 
\hspace{4mm} {\rm for} \hspace{2mm}j{\rm :even }.
\eqend

Then,   the grand-canonical partition function for odd $N$  is given by 
\eqbegin 
Z^{\rm odd}(\tau,\zeta)&=& {\rm Tr}_{\Omega^{\rm odd}_{\rm edge}}
\left( {\rm e}^{  2\pi i\tau(M_E-\frac{n_bc}{24})+2\pi i\zeta Q} \right)
  \nonumber \\
&=& \sum_{j=0}^{k}\sum_{r=0}^{q-1} 
\left[\chi_{j,r}^{\rm even}\chi_{k-j,r}^{\rm odd}+
\chi_{j,r}^{\rm odd}\chi_{k-j,r}^{\rm even}
\right] 
\eqend    
Also the grand-canonical partition function for even  $N$  is given by 
\eqbegin 
Z^{\rm even}(\tau,\zeta)&=& {\rm Tr}_{\Omega^{\rm even}_{\rm edge}}
\left( {\rm e}^{  2\pi i\tau(M_E-\frac{n_bc}{24})+2\pi i\zeta Q} \right)
  \nonumber \\
&=& \sum_{j=0}^{k}\sum_{r=0}^{q-1}\left[(\chi_{j,r}^{\rm even})^2+
(\chi_{j,r}^{\rm odd})^2 \right] 
\eqend 
By introducing  complex conjugate variables to distinguish the two edges, 
we see that the sum of these partition functions  
 $Z^{\rm even}+Z^{\rm odd}$ is  
\eqbegin 
Z^{\rm ann}(\tau, \zeta)=
\sum_{j=0}^{k}\sum_{r=0}^{q-1}|
\chi_{j,r}^{\rm even}+\chi_{k-j,r}^{\rm odd}|^2.  
\label{su2ann}
\eqend
Thus we see that edge excitations of this system 
have $(k+1)q$ sectors. This number  is the same as 
the number of the bulk excitations classified by their braiding property.

Except for  the rational torus part, 
the annulus partition function eq.(\ref{su2ann})  
becomes a $D$ type  $\Gamma(2)$ modular invariants  
 of $\widehat{su(2)}$  in the recent classification by  Gannon \cite{gannon}.  
When $k$ is a multiple of $4$, it is in the regime of 
$ADE$ classification of the $\widehat{su(2)}$ modular 
invariants \cite{ade}. 
This is because the representation 
for the electron is the simple current 
which gives the corresponding 
automorphism of the fusion algebra to 
make $D$-type modular invariants.

\subparagraph{k:odd}
When $k$ is odd, $q$ is a half integer. Let $p$ be $2q=2s+k$. $p$ is 
an odd integer.   
We can  rewrite  the couplings  (\ref{su2coup1})(\ref{su2coup2})  as 
\eqbegin    
V_{j}e^{i\frac{m}{\sqrt{2p}}\varphi}, \hspace{4mm}  m=2n+j \hsf 
n\in \N.  
\label{su2coup}   
\eqend 
The charge sectors  for  the two edges 
$m_1,m_2, \hspace{3mm} m_1,m_2 \in Z$ satisfy 
\eqbegin 
m_1=m_2+Nq.
\label{cc}
\eqend  
When $N$ is even, $\beta_1=\beta_2$ as in (\ref{su2c}).  
$m_1$ and $m_2$ are even or odd
according to  $\beta_1$ is even or odd.  When $\beta_1$ is even, 
by putting $m_1=2m_1', \hspace{2mm}m_2=2m_2', \hspace{2mm} N=2N'$, 
 (\ref{cc}) becomes 
\eqbegin 
m_1'=m_2'+N'p. 
\label{ncc}
\eqend 
When $\beta_1$ is odd, ({\ref{cc}}) reduces to (\ref{ncc})
by putting $m_1-1=2m_1', \hspace{2mm} m_2-1=2m_2'$. 
Thus there are $p$ charge sectors.  

When $N$ is odd, (\ref{cc}) can be rewritten as 
($N=2N'-1$) 
\eqbegin 
m_1=m_2+p+2N'p. 
\label{cc2}
\eqend 
Since $\beta_1=k-\beta_2$ as in (\ref{su2c})  
and  $k$ is odd, $\beta_1$ ($m_1$)and $\beta_2$ ($m_2$)  
have opposite parities.      
When $\beta_1$ is even and $\beta_2$ is odd,  
we can put $m_1=2m_1', \hspace{2mm} m_2=2m_2'-p$. 
Then (\ref{cc2}) again reduces to the form (\ref{ncc}).  
The case that $\beta_1$ is odd is similarly worked out. 
Thus there are also $p$ charge sectors.

Accordingly  we introduce the characters: 
\eqbegin 
\chi_{r/p}(\tau,\zeta)&=& \chi_{r/p}^{\rm even}(\tau,\zeta)+
\chi_{r/p}^{\rm odd}(\tau,\zeta) \nonumber \\ 
&=& 
{{\rm e}^{ -{\pi\over p}{\left(\I\zeta\right)^2\over \I\tau} }\over
  \eta}\ \sum_{m \in Z }
{\rm e}^{ 2\pi i\tau\frac{(mp+r)^2}{2p}+2\pi
  i\zeta(m+\frac{r}{p})}.   
\eqend 

From the consideration above, it is convenient to 
introduce the following characters: 
\eqbegin 
\chi_{j,r}&=&\chi_j^{(k)}\chi_{(2r+p)/2p}^{a}, \hspace{4mm} {\rm for}\hspace{2mm}j{\rm :odd }\\
\chi_{j,r}&=&\chi_{j}^{(k)}\chi_{2r/2p}^{a} 
\hspace{4mm} {\rm for}\hspace{2mm}j{\rm :even }.
\eqend

Then the grand partition function for even  $N$  becomes
\eqbegin 
Z^{\rm even}(\tau,\zeta)&=& {\rm Tr}_{\Omega^{\rm even}_{\rm edge}}
\left( {\rm e}^{ 2\pi i\tau(M_E-\frac{n_bc}{24})+2\pi i\zeta Q} \right)
  \nonumber \\
&=& \sum_{j:\rm even}\sum_{r=0}^{p-1}\left[(\chi_{j,r})^2+
(\chi_{k-j,r})^2 \right]. 
\eqend 
Also the grand-canonical partition function for odd  $N$  becomes  
\eqbegin 
Z^{\rm odd}(\tau,\zeta)&=& {\rm Tr}_{\Omega^{\rm odd}_{\rm edge}}
\left( {\rm e}^{ 2\pi i\tau(M_E-\frac{n_bc}{24})+2\pi i\zeta Q} \right)
  \nonumber \\
&=& \sum_{j: \rm even}\sum_{r=0}^{p-1} 
2\chi_{j,r}\chi_{k-j,r}.  
\eqend    
Then the grand partition function $Z^{\rm even}+Z^{\rm odd}$ becomes 
\eqbegin 
Z^{\rm ann}(\tau,\zeta)=
\sum_{r=0}^{p-1}\sum_{j: {\rm even}}|
\chi_{j,r}+\chi_{k-j,r}|^2.  
\label{su2annkodd}
\eqend
Thus there are $(k+1)q$ sectors of edge excitations as in the even
 $k$ case.   However the Hilbert space of edge excitations is 
decomposed in a different way.

\paragraph{Multiple Boundaries}
We can also deduce the grand partition function for the FQH state on
multiply connected region. 
\subparagraph{k: even}
Next let us consider a spinon-holon FQH 
 state on a region with $n_b$ boundaries. 
In this case, we need at least $n_b-1$ CVOs to express 
the $n_b$ boundaries.  As in the case of an annulus, 
we  end up with an expression, 
\eqbegin
 \iLamda \CVO{\beta_2}{\beta_1}{\alpha_1}
 \iLamda \CVO{\beta_3}{\alpha_1}{\alpha_2} \cdots 
 \iLamda \CVO{\beta_{n_b}}{\alpha_{n_b-2}}{\beta} 
\widetilde{\iPhi}_{\rm disk}
(z_1, \cdots, z_N)
\label{su2kmul}
\eqend 
Let us denote the partition function of edge excitations with
 $\beta=(j,\lamda,a)$ where $j=1,\cdots,k, \hspace{2mm}   
 \lamda=0,\cdots, q-1$, $a={\rm even,odd}$ as $Z_{i,\lamda}^{a,(n_b)}$.
By summing over all the sectors of
 $\beta_{n_b}$, we find the equation satisfied by $Z_{i,\lamda}^{a,(n_b)}$: 
\eqbegin 
 Z_{i,s}^{a_1,(n_b)}=\sum_{jk}\sum_{a_2+a_3=a_1}\sum_{\lamda=0}^{q-1}
 N^{k}_{ij}\chi_{j,\lamda}^{a_2}Z^{a_3,(n_b-1)}_{k,s+\lamda}.
\eqend   
Here $N^{k}_{ij}$ is the fusion rules of $SU(2)_k$ WZW model 
in (\ref{su2fusion}) 
and $Z^{(n_b-1)}$ is the partition function for the remaining 
$(n_b-1)$ boundaries. We can  diagonalize  this equation by 
using   the Verlinde formula \cite{ino} 
\eqbegin 
N_{jk}^{i}&=& \sum_n S_j^{n} \lamda_k^{(n)} S_n^{\dag i}, \\ 
\label{Ver}
\lamda_k^{(n)}&=&{S_k^{n}}/{S^{n}_{0}},  
\eqend
where $S_{i}^{j}$ is the matrix element of modular transformation on 
the Virasoro characters. 
The modular behavior of the characters $\chi^{(k)}_l$ are given by
\cite{gepner,ver}
\eqbegin 
S^{(k)}_{ln}=\left(\frac{2}{k+2}\right)^{1/2}{\rm  sin}\frac{(l+1)(n+1)}{k+2}\pi . 
\label{gep}
\eqend 
By using the modular matrix element, 
we  obtain the partition functions as : 
\eqbegin 
Z^{a,(n_b)}_{l,s}=\sum_{n=0}^{k} 
\left(\frac{2}{k+2}\right)^{1/2}{\rm  sin}
\left(\frac{(l+1)(n+1)}{k+2}\pi\right)\Xi_{n,s}^{a,(n_b)}, 
\eqend
where $\Xi_{n.s}^{a,(n_b)}$ is     
\eqbegin 
\Xi_{n,s}^{a,(n_b)}=\left(
\frac{2}{k+2}\right)^{1/2}
{\rm  sin}\frac{n+1}{k+2}\pi
\sum_{p_2,\cdots,p_n}\sum_{a_{2},\cdots,a_{n_b}}\xi^{a_1}_{n,p_1}\cdots\xi^{a_{n_b}}_{n,p_{n_b}}, \\ 
p_1\equiv s+\sum_{l=2}^{n_b}p_l,  \hspace{3mm} 
(\mbox{mod}\hspace{2mm} 2q),\hspace{4mm} a_1\equiv a+\sum_{l=2}^{n_b}
a_l,  \nonumber 
\eqend 
\eqbegin 
\xi^{a}_{n,p}=
\sum_{j=0}^{k}\left[ \frac{{\rm  sin}\frac{(n+1)(j+1)}{k+2}\pi}
{{\rm  sin}\frac{n+1}{k+2}\pi}\right] \chi_{j,\lamda}^{a}. 
\eqend 
In particular, the grand-canonical partition function 
$Z^{\rm even}_{0,0}+Z^{\rm odd}_{k,0}$ is 
\eqbegin 
Z=
\left(\frac{2}{k+2}\right)^{1/2}\sum_{n=0}^{k} \left[{\rm  sin}
\left(\frac{n+1}{k+2}\pi\right)\left(\Xi_{n,s}^{{\rm even},(n_b)}
+(-1)^{n+1}\Xi_{n,s}^{{\rm odd},(n_b)}\right)
\right].
\eqend

\subparagraph{k:odd}
In this case, we have $p$ different sectors for the charge sector and 
no distinction between even and odd sectors. Thus  we get 
\eqbegin 
Z_{i,s}^{(n_b)}=\sum_{jk}\sum_{\lamda=0}^{2p-1}
 N^{k}_{ij}\chi_{j,\lamda}Z^{(n_b-1)}_{k,s+\lamda}.
\eqend   
Accordingly we get  
\eqbegin 
Z^{(n_b)}_{l,s}=\sum_{n=0}^{k} 
\left(\frac{2}{k+2}\right)^{1/2}{\rm  sin}
\left(\frac{(l+1)(n+1)}{k+2}\pi\right)\Xi_{n,s}^{(n_b)}, 
\eqend
where $\Xi_{n.s}^{(n_b)}$ is     
\eqbegin 
\Xi_{n,s}^{(n_b)}=\left(
\frac{2}{k+2}\right)^{1/2}
{\rm  sin}\frac{n+1}{k+2}\pi
\sum_{p_2,\cdots,p_n}\xi_{n,p_1}\cdots\xi_{n,p_{n_b}}, \\ 
p_1\equiv s+\sum_{l=2}^{n_b}p_l,  \hspace{3mm} 
(\mbox{mod} p), \nonumber 
\eqend 
\eqbegin 
\xi_{n,\lamda}=
\sum_{j=0}^{k}\left[ \frac{{\rm  sin}\frac{(n+1)(j+1)}{k+2}\pi}
{{\rm  sin}\frac{n+1}{k+2}\pi}\right] \chi_{j,\lamda}. 
\eqend 
In particular, the grand-canonical partition function 
$Z_{0,0}+Z_{k,0}$ is 
\eqbegin 
Z=
\left(\frac{2}{k+2}\right)^{1/2}\sum_{n=0}^{k} \left[{\rm  sin}
\left(\frac{n+1}{k+2}\pi\right)\left(\Xi_{n,s}^{(n_b)}
+(-1)^{n+1}\Xi_{n,s}^{(n_b)}\right)
\right]. 
\eqend

The partition function for the Halperin state is   
the $k=1$ case.

\section{Modular Invariance and the Topological Order}
\sectionnumbering

The annulus partition functions 
for paired states (Pfaffian, Haldane-Rezayi, 331) were 
derived in \cite{milo}, but the modular properties of the 
extended characters have not been worked out.  For the Haldane-Rezayi state, 
the modular invariant partition function was discussed in 
\cite{gurarie}, but only for the internal degrees of freedom 
given by the $c=-2$ scalar fermion.
In this section, we investigate the 
full modular behaviors of the characters including all 
the degrees of freedom and verify 
the modular invariance of annulus partition functions for 
the Pfaffian, 331  and spinon-holon states. 
This enable us to   extend the observation of 
Cappelli and Zemba on the relation between 
topological order and modular properties for the 
characters  to these states. 
For the Haldane-Rezayi state, we find a  pathologic 
behavior due to the coupling between 
the internal degrees of freedom and  the charge degrees of freedom.

\subsection{Modular Invariance Conditions}
\label{mic}
In Ref.\cite{cappelli2}
Cappelli and Zemba gave the following extended 
modular invariance conditions 
for the 
annulus partition function of a FQH state:  
\eqbegin 
(S):\qquad Z\left(-{1\over\tau},-{\zeta\over\tau}\right)&=&
Z\left(\tau,\zeta\right)\ .
\label{scond}\\
(T^2):\qquad Z\left(\tau +2,\zeta\right)
&=& Z\left(\tau, \zeta\right)\ .
\label{tcond} \\ 
(U):\qquad Z\left(\tau,\zeta +1\right)&=& Z\left(\tau, \zeta\right)\ .
\label{ucond}\\
(V):\qquad Z\left(\tau,\zeta +\tau\right) &=& Z\left(\tau, \zeta\right)\ .
\label{vcond} 
\eqend 
The physical meanings of these conditions are as follows.  
The edge theory is equivalent to Chern-Simons theory on 
the 3-manifold, $Annulus \times S^{1}$ ($S^{1}$ is the Euclidean time) 
\cite{witten}. The annulus partition function is 
the amplitude in Chern-Simons theory. Then it must be   
invariant under the modular transformation on  the 
modular parameter of  Euclidean space-time or space-space 
torus.  This gives the condition ($S$).    

The condition ($T^{2}$) means that the FQH state is consist of 
electrons and the physical excitations must have integer or half
integer spin.   Also the total charge of the state must be 
integer, which gives the condition ($U$). 
 
The final condition ($V$) is first  considered by Laughlin
\cite{laugh}. It is  a consequence of gauge invariance : 
the addition of a quantum
of flux through the center of the annulus is a symmetry of the
gauge-invariant Hamiltonian but causes a flow of all the quantum states
among themselves (spectral flow). 
The spectral flow can be  simulated by requiring the invariance of 
the partition functions under a shift of the electric potential 
by $\zeta\to\zeta +\tau$.   This shift causes the transport of 
a fractional charge between the two edges and thus determine 
the Hall conductivity.

Among the transformations $S, T^{2}, U$,  and $V$, 
 the transformations $ST^2S$ and $S$ generate the subgroup $\Gamma (2)$
of the modular group $\Gamma=SL(2,\Z)/\Z_2$ of rational
transformations $\tau \to (a\tau +b)/(c\tau +d)$, $a,b,c,d\in \Z$,
which are subjected to the conditions $(a,d)$ odd and $(b,c)$ even.
Thus, $Z$ is invariant under the modular subgroup $\Gamma (2)$ 
\footnote{$S$ and $T^2$ actually generate a slightly larger subgroup
  than 
 $\Gamma(2)$.  See the appendix of Ref.\cite{cappelli2}}.

\subsection{Modular Behavior of U(1) Characters}
As we saw in Sec.2, the coupling of 
chiral boson to internal degrees of freedom are determined by 
the single-valuedness and nonsingularity of wave function in 
 electron's coordinates. 
When we consider the 
grand-canonical ensemble of electrons, we end up  with the 
characters : 
\[\chi_{r/q}^{\rm even}(\tau,\zeta), \hsf
\chi_{r/q}^{\rm odd}(\tau,\zeta),\hsf 
\chi_{(r+1/2)/q}^{\rm even}(\tau,\zeta),\hsf  
\chi_{(r+1/2)/q}^{\rm odd}(\tau,\zeta). \]
Let us determine the modular behavior of these characters. To that
end,  we recall the  character for $q \rightarrow 4q$:
\eqbegin 
\chi_{r/4q}(\tau,\zeta)= 
{{\rm e}^{ -{\pi\over 4q}{\left(\I\zeta\right)^2\over \I\tau} }\over\eta}
\sum_{m \in Z}
{\rm e}^{ 2\pi i\tau\frac{(4qm+r)^2}{8q}+2\pi i\zeta(m+\frac{r}{4q})}.
\label{chara4q}
\eqend 
By comparing (\ref{charaq}-\ref{charaq4}) 
with  (\ref{chara4q}), we see that 
these characters are related as 
\eqbegin 
\chi_{r/q}^{\rm even}(\tau,\zeta)&=&\chi_{2r/4q}(\tau,2\zeta),\label{q4q}\\
\chi_{r/q}^{\rm odd}(\tau,\zeta)&=&\chi_{(2r+2q)/4q}(\tau,2\zeta),\\
\chi_{(r+1/2)/q}^{\rm
  even}(\tau,\zeta)&=&\chi_{(2r+1)/4q}(\tau,2\zeta),\\
\chi_{(r+1/2)/q}^{\rm
  odd}(\tau,\zeta)&=&\chi_{(2r+2q+1)/4q}(\tau,2\zeta).
\label{q4qe}
\eqend

Let us now examine the transformation properties of the characters.   
First, the transformations $T^2,S, U,V$ of the  character 
$\chi_{\lamda/q}$ are  :
\eqbegin 
S:\ \chi_{\lambda/q}\left(-{1\over\tau},-{\zeta\over\tau}\right) & = &
{ {\rm e}^{\ i{\pi\over q}\R {\zeta^2 \over \tau} } \over\sqrt{q}}
\ \sum_{\lambda^\prime=0}^{q-1}\
{\rm e}^{ 2\pi i{\lambda\lambda^\prime\over q} } \
\chi_{\lambda^\prime/q}\left(\tau,\zeta\right)\ ,
\label{chitrS}\\
T^2:\ \chi_{\lambda/q}\left(\tau+2,\zeta\right)\ & = &
{\rm e}^{\ 2\pi i\left({\lambda^2\over q}-{1\over 12} \right) }
\chi_{\lambda/q}\left(\tau,\zeta\right)\ ,
\label{chitrT}
\\
U:\ \chi_{\lambda/q}\left(\tau,\zeta+1\right)\ & = &
{\rm e}^{\ {2\pi i\lambda \over  q}} \
\chi_{\lambda/q}\left(\tau,\zeta\right)\ , 
\label{chitrU}\\
V:\ \chi_{\lambda/q}\left(\tau,\zeta+\tau\right)\ & = &
{\rm e}^{ -{2\pi i\over q}\left(\R\zeta +\R{\tau\over 2} \right) }
\chi_{{(\lambda+1)/q}}\left(\tau,\zeta\right)\ .
\label{chitrV}
\eqend 
These transformations show that the characters $\chi_\lambda/q$ carry a
unitary \rep of the modular group $\Gamma(2)$, which is projective
for $\zeta\neq 0$ (the composition law is verified up to a phase).
Actually, the prefactor is necessary to transform 
$\chi_{\lamda/q}$ by phase under the spectral flow $V$
\cite{cappelli2}.  The same  prefactor also appears in the quantization of
the Chern-Simons theory on the space torus, in the measure
for the inner product  of the wave functions \cite{witten}.

From (\ref{q4q}-\ref{q4qe}), 
the modular behaviors 
of $\chi^{\rm even}_{r/q}, \chi^{\rm odd}_{r/q},
\chi^{\rm even}_{(r+1/2)/q}, 
\chi^{\rm odd}_{(r+1/2)/q}$ under $T^{2}, U,V$ are the same 
with (\ref{chitrT})(\ref{chitrU})(\ref{chitrV}). 
Also the modular behavior under $S$  is determined to be  
\eqbegin 
\chi^{\rm even}_{r/q} &\rightarrow& { {\rm e}^{\ i{\pi\over q}\R
    {\zeta^2 \over \tau} } \over 2\sqrt{q}} \sum_{s=0}^{q-1}
\Biggl\{ {\rm e}^{ \frac{2\pi irs}{q}}(\chi^{\rm even}_{s/q}
+\chi^{\rm odd}_{s/q})+{\rm e}^{ \frac{2\pi ir(s+\frac{1}{2})}{q}}
(\chi^{\rm even}_{(s+1/2)/q}
+\chi^{\rm odd}_{(s+1/2)/q})\Biggr\}, \nonumber \\ 
\label{evchi0}
\\
\chi^{\rm odd}_{r/q} &\rightarrow& { {\rm e}^{\ i{\pi\over q}\R
    {\zeta^2 \over \tau} } \over2\sqrt{q}} \sum_{s=0}^{q-1}
\Biggl\{ {\rm e}^{ \frac{2\pi irs}{q}}(\chi^{\rm even}_{s/q}
+\chi^{\rm odd}_{s/q})-{\rm e}^{ \frac{2\pi ir(s+\frac{1}{2})}{q}}
(\chi^{\rm even}_{(s+1/2)/q}
+\chi^{\rm odd}_{(s+1/2)/q})\Biggr\},
\nonumber \\
\\ 
\chi^{\rm even}_{(r+1/2)/q} &\rightarrow& { {\rm e}^{\ i{\pi\over q}\R
    {\zeta^2 \over \tau} } \over2\sqrt{q}} \sum_{s=0}^{q-1}
\Biggl\{ {\rm e}^{ \frac{2\pi i(r+\frac{1}{2})s}{q}}(\chi^{\rm even}_{s/q}
-\chi^{\rm odd}_{s/q})+{\rm e}^{ \frac{2\pi i(r+\frac{1}{2})(s+\frac{1}{2})}{q}}
(\chi^{\rm even}_{(s+1/2)/q}
-\chi^{\rm odd}_{(s+1/2)/q})\Biggr\},
\nonumber \\
\\ 
\chi^{\rm even}_{(r+1/2)/q} &\rightarrow& { {\rm e}^{\ i{\pi\over q}\R
    {\zeta^2 \over \tau} } \over2\sqrt{q}} \sum_{s=0}^{q-1}
\Biggl\{ {\rm e}^{ \frac{2\pi i(r+\frac{1}{2})s}{q}}(\chi^{\rm even}_{s/q}
-\chi^{\rm odd}_{s/q})-{\rm e}^{ \frac{2\pi i(r+\frac{1}{2})(s+\frac{1}{2})}{q}}(
\chi^{\rm even}_{(s+1/2)/q}
-\chi^{\rm odd}_{(s+1/2)/q})\Biggr\}.  \nonumber \\ 
\label{evchi}
\eqend 

\subsection{Spinon-Holon States} 
Let us examine the modular behavior of characters for spinon-holon
states.  
\subparagraph{k: even}
From (\ref{su2ann}), there are $(k+1)q$ sectors:
\eqbegin 
\chi^{\rm sph}_{j,r}=\chi_{j,r}^{\rm even}+\chi_{k-j,r}^{\rm odd}, \hsf
j=0,\cdots, k,\hspace{2mm} r=0,\cdots,q-1. 
\eqend 
Combining the modular behaviors  in (\ref{gep}) and  
(\ref{evchi0}-\ref{evchi}), we  get the 
modular behavior of $\chi^{\rm sph}_{j,r}$. 
When $j$ is even, we have 
\eqbegin 
\chi_{j,r}^{\rm even} &\rightarrow& 
{ {\rm e}^{\ i{\pi\over q}\R
    {\zeta^2 \over \tau} } \over2\sqrt{q}} 
\sum_{n=0}^{k} \sum_{s=0}^{q-1}
S^{(k)}_{jn} \Biggl\{{\rm e}^{ \frac{2\pi irs}{q}}\left(\chi_{n,s}^{\rm even} 
+\chi_{n,s}^{\rm odd}\right)+ 
{\rm e}^{ \frac{2\pi ir(s+\frac{1}{2})}{q}}\left(\chi_{n,s+1/2}^{\rm even} 
+\chi_{n,s+1/2}^{\rm odd}\right)\Biggr\}, \nonumber 
\\
\\ 
\chi_{j,r}^{\rm odd} &\rightarrow& {{\rm e}^{\ i{\pi\over q}\R
    {\zeta^2 \over \tau} } \over2\sqrt{q}} 
\sum_{n=0}^{k} \sum_{s=0}^{q-1}
S^{(k)}_{jn} \Biggl\{{\rm e}^{ \frac{2\pi irs}{q}}\left(\chi_{n,s}^{\rm even} 
+\chi_{n,s}^{\rm odd}\right)-
{\rm e}^{ \frac{2\pi ir(s+\frac{1}{2})}{q}}\left(\chi_{n,s+1/2}^{\rm even} 
+\chi_{n,s+1/2}^{\rm odd}\right)\Biggr\}, \nonumber 
\\
\eqend 
When $j$ is odd, we have 
\eqbegin  
\chi_{j,r}^{\rm even} &\rightarrow& { {\rm e}^{\ i{\pi\over q}\R
    {\zeta^2 \over \tau} } \over2\sqrt{q}} 
\sum_{n=0}^{k} \sum_{s=0}^{q-1}
S^{(k)}_{jn} \Biggl\{{\rm e}^{ \frac{2\pi i(r+\frac{1}{2})s}{q}}\left(\chi_{n,s}^{\rm even} 
-\chi_{n,s}^{\rm odd}\right)+ {\rm e}^{ \frac{2\pi i(r+\frac{1}{2})
(s+\frac{1}{2})}{q}}\left(\chi_{n,s+1/2}^{\rm even} 
-\chi_{n,s+1/2}^{\rm odd}\right)\Biggr\}, \nonumber 
\\
\\ 
\chi_{j,r}^{\rm odd} &\rightarrow& { {\rm e}^{\ i{\pi\over q}\R
    {\zeta^2 \over \tau} } \over2\sqrt{q}} 
\sum_{n=0}^{k} \sum_{s=0}^{q-1}
S^{(k)}_{jn} \Biggl\{ {\rm e}^{ \frac{2\pi i(r+\frac{1}{2})s}{q}}
\left(\chi_{n,s}^{\rm even} 
-\chi_{n,s}^{\rm odd}\right)  
-{\rm e}^{ \frac{2\pi ir(s+\frac{1}{2})}{q}}\left(\chi_{n,s+1/2}^{\rm even} 
-\chi_{n,s+1/2}^{\rm odd}\right)\Biggr\}. \nonumber 
\\ 
\eqend 
By noticing 
\eqbegin 
S^{(k)}_{(k-l)n} = (-1)^{n}S^{(k)}_{ln}
\label{su2r}
\eqend 
we see that  the modular behavior of $\chi^{\rm sph}_{j,r}$ is 
given by ,
\eqbegin 
\widetilde{S}: \chi^{\rm sph}_{j,r} &\rightarrow&
\frac{{\rm e}^{\ i{\pi\over q}\R{\zeta^2 \over \tau}}}{\sqrt{q}} 
\Bigl\{\sum_{n :\rm even} \sum_{s=0}^{q-1} 
S^{(k)}_{jn}{\rm e}^{ \frac{2\pi irs}{q}}\chi^{\rm sph}_{n,s}
+\sum_{n :\rm odd}\sum_{s=0}^{q-1} 
S^{(k)}_{jn}{\rm e}^{ \frac{2\pi ir(s+\frac{1}{2})}{q}}\chi^{\rm
  sph}_{n,s}\Bigr\}, 
\hspace{3mm}{\rm for} \hspace{2mm}j \hspace{2mm}{\rm even},  
\\ 
\widetilde{S}:  \chi^{\rm sph}_{j,r} &\rightarrow&
\frac{{\rm e}^{\ i{\pi\over q}\R{\zeta^2 \over \tau}}}{\sqrt{q}} 
\Bigl\{\sum_{n :\rm even} \sum_{s=0}^{q-1} 
S^{(k)}_{jn}{\rm e}^{ \frac{2\pi i(r+\frac{1}{2})s}{q}}\chi^{\rm sph}_{n,s}
+\sum_{n :\rm odd}\sum_{s=0}^{q-1} 
S^{(k)}_{jn}{\rm e}^{ \frac{2\pi i(r+\frac{1}{2})(s+\frac{1}{2})}{q}}\chi^{\rm
  sph}_{n,s}\Bigr\}, \hspace{3mm}{\rm for}\hspace{2mm} j\hspace{2mm}
{\rm odd}. 
  \nonumber \\
\label{keven}
\eqend 
Thus $\chi^{\rm sph}_{j,r}$ for $j$ :even, odd 
form an irreducible  representation of the modular group.  
It is readily seen that the annulus partition function (\ref{su2ann}) 
is invariant under $(S)$.  The conditions 
$(T^2)$, $(U)$, and $(V)$ are  verified
similarly.   
We also see that $\widetilde{S}$ reproduces the 
correct fusion rules of the representations in this model 
through the Verlinde formula (\ref{ver}).

\subparagraph{k:odd}
(\ref{su2annkodd}) can be rewritten as 
\eqbegin 
Z^{\rm ann}(\tau, \zeta)=\frac{1}{2}
\sum_{r=0}^{p-1}\sum_{j=0}^{k}|
\chi_{j,r}+\chi_{k-j,r}|^2 
\label{su2odd2}
\eqend 
Thus the sectors of edge excitations are 
\eqbegin  
\chi^{\rm sph}_{j,r}
=\chi_{j,r}+\chi_{k-j,r}, \hsf
j=0,\cdots, k,\hspace{2mm} r=0,\cdots,p-1. 
\eqend 
In this case,  $\chi_{j,r}$ transforms as
\eqbegin 
\chi_{j,r} &\rightarrow& { {\rm e}^{\ i{\pi\over q}\R
    {\zeta^2 \over \tau} } \over2\sqrt{q}} 
\sum_{s=0}^{2p-1}  
\sum_{n=0}^{k} {\rm e}^{ {2\pi i rs \over p}} S^{(k)}_{jn} \chi_{n,s/2}  \hsf
{\rm for} \hspace{2mm}j:{\rm even} \\ 
\chi_{j,r} &\rightarrow& { {\rm e}^{\ i{\pi\over q}\R
    {\zeta^2 \over \tau} } \over2\sqrt{q}} 
\sum_{s=0}^{2p-1}   
\sum_{n=0}^{k} {\rm e}^{ {2\pi i rs \over p}}{\rm e}^{ i\pi s}
S^{(k)}_{jn} \chi_{n,s/2}  \hsf {\rm for} \hspace{2mm}j:{\rm odd}
\eqend  
Therefore, from (\ref{su2r}), 
$\chi^{\rm sph}_{j,r}$ transforms as 
\eqbegin 
\chi^{\rm sph}_{j,r} \rightarrow 
{ {\rm e}^{\ i{\pi\over q}\R
    {\zeta^2 \over \tau} } \over \sqrt{q}}\sum_{s=0}^{p-1}\sum_{n=0}^{k}
{\rm e}^{{2 \pi i rs \over p}}S^{(k)}_{jn}\chi^{\rm sph}_{n,s}.   
\label{kodd}
\eqend 
We see that the state realizes a different representation from the 
even $k$ case (\ref{keven}).  
Obviously the annulus partition function for odd $k$ (\ref{su2odd2})
is invariant under this transformation. The conditions 
$(T^2)$, $(U)$, and $(V)$ are verified
similarly.    
The fusion rules are also reproduced from the Verlinde formula.

The degeneracy of the state on a genus $g$ Riemann surface is 
$(k+1)^g q^g$ for both even and odd $k$. However the structures 
of the topological order are different as we see from (\ref{keven}) 
and (\ref{kodd}).   

\subsection{Paired States}
In this section, 
we examine  the modular properties of characters  for paired states. 
We first present the analysis for the Pfaffian state \cite{moore} 
in detail. 
\subsubsection{The Pfaffian state}
The annulus partition function for the Pfaffian state at 
$\nu=\frac{1}{q}$ is  given by \cite{milo} 
\eqbegin 
Z^{\rm ann}_{\rm Pf}&=&\sum_{r=0}^{q-1}\biggl\{ |\chi_{1,r}^{\rm even}+
\chi_{\psi,r}^{\rm odd}|^2+ 
|\chi_{\psi,r}^{\rm even}+
\chi_{1,r}^{\rm odd}|^2+
|\chi^{\rm even}_{\sigma,r}+\chi^{\rm odd}_{\sigma,r}|^2  \biggr\} 
\label{pfan}
\\ 
\chi_{i,r}^{a}&=&\chi_i^{\rm MW}\chi_{r/q}^{a}, \hspace{4mm} i=1,\psi,  \\
\chi_{\sigma,r+\frac{1}{2}}^{a}&=&\chi_{\sigma}^{\rm MW}\chi_{(r+1/2)/q}^{a}, 
\eqend
where $\chi^{\rm MW}_{1},\chi^{\rm MW}_{\psi},\chi^{\rm MW}_{\sigma}$ 
are the characters for the Ising model: 
\eqbegin 
\chi^{\rm MW}_1(\tau)&=&\frac{1}{2}\omega^{-\frac{1}{48}}
\left(\prod_{0}^{\infty}(1+\omega^{n+\frac{1}{2}}) +
 \prod_{0}^{\infty}(1-\omega^{n+\frac{1}{2}}) 
 \right) \\ 
 &=&\frac{1}{2}\sqrt{\frac{\theta_3(\tau)}{\eta(\tau)}}+\frac{1}{2}\sqrt{\frac{\theta_4(\tau)}{\eta(\tau)}}, \\
\chi^{\rm MW}_{\sigma}(\tau)&=&
 \omega^{\frac{1}{24}}\prod_{1}^{\infty}(1+\omega^n) \\ 
&=&\sqrt{\frac{\theta_2(\tau)}{2\eta(\tau)}}, \\  
 \chi^{\rm MW}_{\psi}(\tau)&=&
 \frac{1}{2}\omega^{-\frac{1}{48}}\left(\prod_{0}^{\infty}(1+\omega^{n+\frac{1}{2}}) -
 \prod_{0}^{\infty}(1-\omega^{n+\frac{1}{2}}) 
 \right)\\ 
 &=&
 \frac{1}{2}\sqrt{\frac{\theta_3(\tau)}{\eta(\tau)}}
 -\frac{1}{2}\sqrt{\frac{\theta_4(\tau)}{\eta(\tau)}}.  
\eqend  
Here $\theta_2, \theta_3, \theta_4$ are Jacobi theta functions. 
The matrix elements of modular transformation for the Virasoro 
characters 
$\chi^{\rm MW}_1,\chi^{\rm MW}_{\sigma},\chi^{\rm MW}_{\psi}$ 
 follow from the transformation property of the Jacobi 
 $\theta$ functions as  
\eqbegin 
S_{\rm Ising}=\frac{1}{2}\left(
\begin{array}{ccc}
1 & \sqrt{2} &   1\\
\sqrt{2} & 0 & -\sqrt{2}         \\
1 & -\sqrt{2} &  1     \\
\end{array} 
\right).
\label{isingS}
\eqend

From (\ref{pfan}),   
we see that there are $3q$ sectors in the edge excitations of the 
Pfaffian state: 
\eqbegin 
\chi^{\rm Pf}_{1,r}&=&\chi_{1,r}^{\rm even}+
\chi_{\psi,r}^{\rm odd}\\
\chi^{\rm Pf}_{\psi,r}&=&\chi_{\psi,r}^{\rm even}+
\chi_{1,r}^{\rm odd} \\ 
 \chi^{\rm Pf}_{\sigma,r+1/2}&=&\chi^{\rm even}_{\sigma,r+1/2}+\chi^{\rm
   odd}_{\sigma,r+1/2} 
\eqend

From (\ref{isingS}) and (\ref{evchi}), we get the 
modular behaviors  
\eqbegin 
\chi_{1,r}^{\rm even} &\rightarrow& 
{ {\rm e}^{\ i{\pi\over q}\R{\zeta^2 \over \tau} } \over 2\sqrt{q}}
\sum_{s=0}^{q-1}  \Biggl{\{} 
 \frac{1}{2}\biggl({\rm e}^{ \frac{2 \pi irs}{q}}
\bigl(\chi^{\rm even}_{1,s}+\chi^{\rm odd}_{1,s}\bigr)+
{\rm e}^{ \frac{2 \pi ir(s+\frac{1}{2})}{q}}
\bigl(\chi^{\rm even}_{1,s+1/2}+\chi^{\rm odd}_{1,s+1/2}\bigr)
\biggr) \nonumber \\ 
&+& \frac{\sqrt{2}}{2}\biggl({\rm e}^{ \frac{2\pi irs}{q}}\bigl(
\chi^{\rm even}_{\sigma,s}+
\chi^{\rm odd}_{\sigma,s}\bigr)+{\rm e}^{ \frac{2\pi ir(s+\frac{1}{2})}{q}}
\bigl(\chi^{\rm even}_{\sigma,s+1/2}+\chi^{\rm
    odd}_{\sigma,s+1/2}\bigr)
\biggr) \nonumber \\ 
&+& \frac{1}{2}\biggl(
{\rm e}^{ \frac{2 \pi irs}{q}}
\bigl(\chi^{\rm even}_{\psi,s}+\chi^{\rm odd}_{\psi,s}\bigr)+
{\rm e}^{ \frac{2 \pi ir(s+\frac{1}{2})}{q}}
\bigl(\chi^{\rm even}_{\psi,s+1/2}+\chi^{\rm odd}_{\psi,s+1/2}\bigr)
\biggr) \Biggr\},  
\nonumber \\
\\ 
\chi_{1,r}^{\rm odd} &\rightarrow& { {\rm e}^{\ i{\pi\over q}\R
    {\zeta^2 \over \tau} } \over 2\sqrt{q}}
\sum_{s=0}^{q-1} \Biggl\{ \frac{1}{2}\biggl(
{\rm e}^{ \frac{2 \pi irs}{q}}
\bigl(\chi^{\rm even}_{1,s}+\chi^{\rm odd}_{1,s}\bigr)-
{\rm e}^{ \frac{2 \pi ir(s+\frac{1}{2})}{q}}
\bigl(\chi^{\rm even}_{1,s+1/2}+\chi^{\rm odd}_{1,s+1/2}\bigr)
\biggr) \nonumber \\ 
&+& \frac{\sqrt{2}}{2}\biggl({\rm e}^{ \frac{2\pi irs}{q}}\bigl(
\chi^{\rm even}_{\sigma,s}+
\chi^{\rm odd}_{\sigma,s}\bigr)-{\rm e}^{ \frac{2\pi ir(s+\frac{1}{2})}{q}}
\bigl(\chi^{\rm even}_{\sigma,s+1/2}+\chi^{\rm odd}_{\sigma,s+1/2}\bigr) 
\biggr) \nonumber \\
&+&\frac{1}{2} \biggl(
{\rm e}^{ \frac{2 \pi irs}{q}}
\bigl(\chi^{\rm even}_{\psi,s}+\chi^{\rm odd}_{\psi,s}\bigr)-
{\rm e}^{ \frac{2 \pi ir(s+\frac{1}{2})}{q}}
\bigl(\chi^{\rm even}_{\psi,s+1/2}+\chi^{\rm odd}_{\psi,s+1/2}\bigr)
\biggr) \Biggr\}, \nonumber
\\
\\
\chi_{1,r}^{\rm even} &\rightarrow&  { {\rm e}^{\ i{\pi\over q}\R
    {\zeta^2 \over \tau} } \over 2\sqrt{q}}
\sum_{s=0}^{q-1}  \Biggl\{ \frac{1}{2}\biggl(
{\rm e}^{ \frac{2 \pi irs}{q}}
\bigl(\chi^{\rm even}_{1,s}+\chi^{\rm odd}_{1,s}\bigr)+
{\rm e}^{ \frac{2 \pi ir(s+\frac{1}{2})}{q}}
\bigl(\chi^{\rm even}_{1,s+1/2}+\chi^{\rm odd}_{1,s+1/2}\bigr)
\biggr) \nonumber \\ 
&-& \frac{\sqrt{2}}{2}\biggl({\rm e}^{ \frac{2\pi irs}{q}}\bigl(
\chi^{\rm even}_{\sigma,s}+
\chi^{\rm odd}_{\sigma,s}\bigr)+{\rm e}^{ \frac{2\pi ir(s+\frac{1}{2})}{q}}
\bigl(\chi^{\rm even}_{\sigma,s+1/2}+\chi^{\rm
    odd}_{\sigma,s+1/2}\bigr)
\biggr) \nonumber \\ 
&+& \frac{1}{2}\biggl(
{\rm e}^{ \frac{2 \pi irs}{q}}
\bigl(\chi^{\rm even}_{\psi,s}+\chi^{\rm odd}_{\psi,s}\bigr)+
{\rm e}^{ \frac{2 \pi ir(s+\frac{1}{2})}{q}}
\bigl(\chi^{\rm even}_{\psi,s+1/2}+\chi^{\rm odd}_{\psi,s+1/2}\bigr)
\biggr) \Biggr\},  \nonumber
\\
\\
\chi_{1,r}^{\rm odd} &\rightarrow&  { {\rm e}^{\ i{\pi\over q}\R
    {\zeta^2 \over \tau} } \over 2\sqrt{q}}
\sum_{s=0}^{q-1}  \Biggl\{\frac{1}{2}\biggl(
{\rm e}^{ \frac{2 \pi irs}{q}}
\bigl(\chi^{\rm even}_{1,s}+\chi^{\rm odd}_{1,s}\bigr)-
{\rm e}^{ \frac{2 \pi ir(s+\frac{1}{2})}{q}}
\bigl(\chi^{\rm even}_{1,s+1/2}+\chi^{\rm odd}_{1,s+1/2}\bigr)
\biggr) \nonumber \\ 
&-& \frac{\sqrt{2}}{2}\biggl({\rm e}^{ \frac{2\pi irs}{q}}\bigl(
\chi^{\rm even}_{\sigma,s}+
\chi^{\rm odd}_{\sigma,s}\bigr)-{\rm e}^{ \frac{2\pi ir(s+\frac{1}{2})}{q}}
\bigl(\chi^{\rm even}_{\sigma,s+1/2}+\chi^{\rm odd}_{\sigma,s+1/2}\bigr) 
\biggr) \nonumber \\
&+&\frac{1}{2}\biggl(
{\rm e}^{ \frac{2 \pi irs}{q}}
\bigl(\chi^{\rm even}_{\psi,s}+\chi^{\rm odd}_{\psi,s}\bigr)-
{\rm e}^{ \frac{2 \pi ir(s+\frac{1}{2})}{q}}
\bigl(\chi^{\rm even}_{\psi,s+1/2}+\chi^{\rm odd}_{\psi,s+1/2}\bigr)
\biggr) \Biggr\},  \nonumber
\\
\\
\chi_{\sigma,r+1/2}^{\rm even} &\rightarrow& { {\rm e}^{\ i{\pi\over q}\R
    {\zeta^2 \over \tau} } \over 2\sqrt{q}} 
\sum_{s=0}^{q-1}  \Bigg\{ \frac{\sqrt{2}}{2}\biggl(
{\rm e}^{ \frac{2 \pi i(r+\frac{1}{2})s}{q}}
\bigl(\chi^{\rm even}_{1,s}-\chi^{\rm odd}_{1,s}\bigr)+
{\rm e}^{ \frac{2 \pi i(r+\frac{1}{2})(s+\frac{1}{2})}{q}}
\bigl(\chi^{\rm even}_{1,s+1/2}-\chi^{\rm odd}_{1,s+1/2}\bigr)
\biggr) \nonumber \\ 
&-& \frac{\sqrt{2}}{2}\biggl(
{\rm e}^{ \frac{2 \pi i(r+\frac{1}{2})s}{q}}
\bigl(\chi^{\rm even}_{\psi,s}-\chi^{\rm odd}_{\psi,s}\bigr)+
{\rm e}^{ \frac{2 \pi i(r+\frac{1}{2})(s+\frac{1}{2})}{q}}
\bigl(\chi^{\rm even}_{\psi,s+1/2}-\chi^{\rm odd}_{\psi,s+1/2}\bigr)
\biggr) \Biggr\},  \nonumber
\\
\\
\chi_{\sigma,r+1/2}^{\rm even} &\rightarrow& { {\rm e}^{\ i{\pi\over q}\R
    {\zeta^2 \over \tau} } \over 2\sqrt{q}} 
\sum_{s=0}^{q-1}  \Bigg\{ \frac{\sqrt{2}}{2}\biggl(
{\rm e}^{ \frac{2 \pi i(r+\frac{1}{2})s}{q}}
\bigl(\chi^{\rm even}_{1,s}-\chi^{\rm odd}_{1,s}\bigr)-
{\rm e}^{ \frac{2 \pi i(r+\frac{1}{2})(s+\frac{1}{2})}{q}}
\bigl(\chi^{\rm even}_{1,s+1/2}-\chi^{\rm odd}_{1,s+1/2}\bigr)
\biggr) \nonumber \\ 
&-& \frac{\sqrt{2}}{2}\biggl(
{\rm e}^{ \frac{2 \pi i(r+\frac{1}{2})s}{q}}
\bigl(\chi^{\rm even}_{\psi,s}-\chi^{\rm odd}_{\psi,s}\bigr)-
{\rm e}^{ \frac{2 \pi i(r+\frac{1}{2})(s+\frac{1}{2})}{q}}
\bigl(\chi^{\rm even}_{\psi,s+1/2}-\chi^{\rm odd}_{\psi,s+1/2}\bigr)
\biggr) \Biggr\}. \nonumber
\\
\eqend

By using these formulas, 
the modular behavior of  each sector in (\ref{pfan}) is 
deduced as 
\eqbegin 
\widetilde{S}:\chi^{\rm Pf}_{1,r} &\rightarrow&  { {\rm e}^{\ i{\pi\over q}\R
    {\zeta^2 \over \tau} } \over 2\sqrt{q}}
\sum_{s=0}^{q-1} 
\left({\rm e}^{ \frac{2 \pi irs}{q}}
\left(\chi_{1,r}^{\rm Pf}+\chi_{\psi,r}^{\rm Pf}\right)
+\sqrt{2} {\rm e}^{ \frac{2 \pi ir(s+\frac{1}{2})}{q}}
\chi_{\sigma,s+\frac{1}{2}}^{\rm Pf} \right), \\
\widetilde{S}:\chi_{\psi,r}^{\rm Pf}  &\rightarrow& 
{ {\rm e}^{\ i{\pi\over q} \R
    {\zeta^2 \over \tau} } \over 2\sqrt{q}}\sum_{s=0}^{q-1}
\left({\rm e}^{ \frac{2 \pi irs}{q}}
\left(\chi_{1,r}^{\rm Pf}+\chi_{\psi,r}^{\rm Pf}\right)
- \sqrt{2} {\rm e}^{ \frac{2 \pi ir(s+\frac{1}{2})}{q}}
\chi_{\sigma,s+\frac{1}{2}}^{\rm Pf} \right),\\
\widetilde{S}:\chi^{\rm Pf}_{\sigma,r+1/2} &\rightarrow& 
{ \sqrt{2}{\rm e}^{\ i{\pi\over q}\R
    {\zeta^2 \over \tau} } \over 2\sqrt{q}}
\sum_{s=0}^{q-1} 
{\rm e}^{ \frac{2 \pi i(r+\frac{1}{2})s}{q}}
\left(\chi_{1,r}^{\rm Pf}-\chi_{\psi,r}^{\rm Pf}\right). 
\eqend
Thus $\chi^{\rm Pf}_{i,r}$ form an irreducible representation 
of the modular group.  
It is readily seen that the annulus partition function (\ref{pfan}) 
is invariant under $(S)$.

Let us recall the fusion rules formed by quasiparticles in the
Pfaffian state. They are 
\eqbegin 
1e^{i\frac{r}{\sqrt{q}}}\times 1e^{i\frac{s}{\sqrt{q}}}&=& 
1e^{i\frac{r+s}{\sqrt{q}}}, \\
1e^{i\frac{r}{\sqrt{q}}}\times \psi e^{i\frac{s}{\sqrt{q}}}&=& 
\psi e^{i\frac{r+s}{\sqrt{q}}}, \\
1e^{i\frac{r}{\sqrt{q}}}\times \sigma e^{i\frac{s+1/2}{\sqrt{q}}}&=& 
\sigma e^{i\frac{r+s+1/2}{\sqrt{q}}}, \\
\psi e^{i\frac{r}{\sqrt{q}}}\times \psi e^{i\frac{s}{\sqrt{q}}}&=& 
1 e^{i\frac{r+s}{\sqrt{q}}}, \\
\psi e^{i\frac{r}{\sqrt{q}}}\times \sigma e^{i\frac{s+1/2}{\sqrt{q}}}&=& 
\sigma e^{i\frac{r+s+1/2}{\sqrt{q}}}, \\
\sigma e^{i\frac{r+1/2}{\sqrt{q}}}\times \sigma e^{i\frac{s+1/2}{\sqrt{q}}}&=& 
1 e^{i\frac{r+s}{\sqrt{q}}}+\psi e^{i\frac{r+s+1}{\sqrt{q}}}. 
\eqend
If we denote these fusion rules by $\widetilde{N^{i}_{jk}}$, it is
confirmed the Verlinde formula for the Pfaffian state 
\eqbegin 
\widetilde{N}_{jk}^{i}&=& \sum_n \widetilde{S}_j^{n} \widetilde{\lamda}_k^{(n)} \widetilde{S}_n^{\dag i}, \\ 
\label{ver}
\widetilde{\lamda}_k^{(n)}&=&\widetilde{S}_k^{n}/\widetilde{S}^{n}_{0}.
\eqend 
This formula can be used to compute the topological order of 
the Pfaffian state as Cappelli and Zemba do for the abelian cases
in \cite{cappelli2}.  For example, the degeneracy of the ground state 
of the Pfaffian state at $\nu=1/q$ on a genus $g$ Riemann surface 
is given by $3^{g}q^{g}$. 

\subsubsection{The Haldane-Rezayi state}
The annulus grand-canonical partition function for the 
Haldane-Rezayi state is given by \cite{milo}
\eqbegin 
Z_{\rm HR}^{\rm ann}=\sum_{r=0}^{q-1} \biggl\{ 2|\chi_{1,r+1/2}+
\chi_{\psi,r+1/2}|^2&+&|\chi^{\rm even}_{\sigma,r}
+\chi^{\rm odd}_{\widetilde{\sigma},r}|^2 
+|\chi^{\rm odd}_{\sigma,r}+\chi^{\rm even}_{\widetilde{\sigma},r}|^2
\biggr\} 
\label{ZHR}\\
\chi_{i,r}&=&\chi_i^{c=-2}\chi_{r/q}, \hspace{4mm} i=1,\psi,
\\
\chi^{a}_{\sigma,r+1/2}&=&\chi_{\sigma}^{c=-2}\chi^{a}_{(r+1/2)/q}, \hspace{3mm}a={\rm even},{\rm odd}
\label{anHR}
\eqend 
 where $\chi^{c=-2}_{1},\chi^{c=-2}_{\psi},\chi^{c=-2}_{\sigma},
\chi^{c=-2}_{\widetilde{\sigma}}$ are
the characters for the ones of $c=-2$ scalar fermion : 
\eqbegin 
\chi_{1}^{c=-2}
&=&
 \frac{1}{2}\omega^{\frac{1}{12}}\left[\prod_{1}^{\infty}(1+\omega^n)^2 +
\prod_{1}^{\infty}(1-\omega^n)^2 \right] \\ 
&=&\frac{1}{2}
\left(\frac{\theta_2(\tau)}{2\eta(\tau)}+\eta(\tau)^2\right)
\\
\chi_{\psi}^{c=-2} &=&
\frac{1}{2}\omega^{\frac{1}{12}}\left[\prod_{1}^{\infty}(1+\omega^n)^2
- \prod_{1}^{\infty}(1-\omega^n)^2 \right] \\ 
&=&\frac{1}{2}
\left(\frac{\theta_2(\tau)}{2\eta(\tau)}-\eta(\tau)^2\right),  
\\
\chi^{c=-2}_{\sigma} 
&=&\frac{1}{2}\omega^{-\frac{1}{24}}
\left(\prod_{0}^{\infty}(1+\omega^{n+\frac{1}{2}})^2 +
 \prod_{0}^{\infty}(1-\omega^{n+\frac{1}{2}})^2
 \right) \\ 
 &=&\frac{1}{2}\left(\frac{\theta_3(\tau)}{\eta(\tau)}
+\frac{\theta_4(\tau)}{\eta(\tau)}\right), 
\\ 
\chi^{c=-2}_{\widetilde{\sigma}}
&=&
 \frac{1}{2}\omega^{-\frac{1}{24}}\left(\prod_{0}^{\infty}(1+\omega^{n+\frac{1}{2}})^2 -
 \prod_{0}^{\infty}(1-\omega^{n+\frac{1}{2}})^2 
 \right)\\ 
 &=&
 \frac{1}{2}\left(\frac{\theta_3(\tau)}{\eta(\tau)}
 -\frac{\theta_4(\tau)}{\eta(\tau)}\right). 
\eqend
The factor $2$ in (\ref{ZHR}) is from the zero modes of the scalar fermions. 
As noted in \cite{flohr,gaber},the behaviors of    
$\chi^{c=-2}_1$ and $\chi^{c=-2}_{\psi}$ under $S$ involve 
the coefficients which are themselves functions of $\tau$ 
(from $\eta^{2}$ terms).  
However the extended characters which appear in (\ref{anHR}) are 
\eqbegin 
\chi^{\rm HR}_{\Psi,r}&=&\sqrt{2}(\chi_{1,r}+\chi_{\psi,r})\equiv 
\sqrt{2}\chi^{c=-2}_{\Psi}\chi_{r/q}, \hspace{3mm}\chi^{c=-2}_{\Psi}=
\chi^{c=-2}_1+\chi^{c=-2}_{\psi}, \label{ec1}\\  
\chi^{\rm HR}_{\sigma,r+1/2}&=&\chi^{\rm even}_{\sigma,r+1/2}+
\chi^{\rm odd}_{\widetilde{\sigma},r+1/2}, \label{ec2} \\
\chi^{\rm HR}_{\widetilde{\sigma},r+1/2}&=&\chi^{\rm even}_{\sigma,r+1/2}+
\chi^{\rm odd}_{\widetilde{\sigma},r+1/2}. \label{ec3}   
\eqend 
Thus $\chi^{c=-2}_1$ and $\chi^{c=-2}_{\psi}$ only  through 
the form   
$\chi^{c=-2}_{\Psi}=\chi^{c=-2}_1+\chi^{c=-2}_{\psi}$ and 
the $\tau$ dependence of the modular matrix elements disappears.   

The extended characters transform under $S$ as 
\eqbegin  
\widetilde{S}:  \chi^{\rm HR}_{\Psi} &\rightarrow&  
{ \sqrt{2}{\rm e}^{\ i{\pi\over q}\R
    {\zeta^2 \over \tau} } \over 2\sqrt{q}}
\sum_{s=0}^{q-1} 
{\rm e}^{ \frac{2 \pi irs}{q}}
\bigl(\chi^{c=-2}_{\sigma}\chi_{s/q}-
\chi^{c=-2}_{\widetilde{\sigma}}\chi_{s/q}\bigr)
\\
\widetilde{S}:  \chi^{\rm HR}_{\sigma,r+1/2} &\rightarrow& 
 { {\rm e}^{\ i{\pi\over q}\R
    {\zeta^2 \over \tau} } \over 2\sqrt{q}}
\sum_{s=0}^{q-1} 
\Biggl\{{\rm e}^{ \frac{2 \pi i(r+\frac{1}{2})s}{q}}
\bigl(\chi^{c=-2}_{\sigma}+\chi^{c=-2}_{\widetilde{\sigma}}\bigr)
\bigl(\chi^{\rm even}_{s/q}-\chi^{\rm odd}_{s/q}\bigr)\nonumber \\
&+&\sqrt{2} {\rm e}^{ \frac{2 \pi i(r+\frac{1}{2})(s+\frac{1}{2})}{q}}
\chi^{c=-2}_{\Psi}
\bigl(\chi^{\rm even}_{(s+1/2)/q}-\chi^{\rm odd}_{(s+1/2)/q}\bigr) 
\Biggr\}  
\\ 
\widetilde{S}: \chi^{\rm HR}_{\widetilde{\sigma},r+1/2} &\rightarrow& 
 { {\rm e}^{\ i{\pi\over q}\R
    {\zeta^2 \over \tau} } \over 2\sqrt{q}}
\sum_{s=0}^{q-1} 
\Biggl\{{\rm e}^{ \frac{2 \pi i(r+\frac{1}{2})s}{q}}
\bigl(\chi^{c=-2}_{\sigma}+\chi^{c=-2}_{\widetilde{\sigma}}\bigr)
\bigl(\chi^{\rm even}_{s/q}-\chi^{\rm odd}_{s/q}\bigr) \nonumber \\
&-&\sqrt{2} {\rm e}^{ \frac{2 \pi i(r+\frac{1}{2})(s+\frac{1}{2})}{q}}
\chi^{c=-2}_{\Psi}
\bigl(\chi^{\rm even}_{(s+1/2)/q}-\chi^{\rm odd}_{(s+1/2)/q}\bigr) 
\Biggr\}. 
\eqend 
The problem here is that the extended characters 
$\chi^{\rm HR}_{\Psi,r}, 
\chi^{\rm HR}_{\sigma,r+1/2}, \chi^{\rm HR}_{\widetilde{\sigma},r+1/2}$
are not closed under the modular transformation. Thus 
they do not form a representation of the modular group.  
Accordingly $Z^{\rm HR}_{\rm ann}$ does not satisfy the 
modular invariance conditions. 

One method to cure this trouble is to map the edge theory by 
the  map  of $c=-2$ scalar 
fermion to  $c=1$ Dirac fermion as in 
\cite{flohr,gurus,gurarie}. 
A simple way to accomplish it is to  insert  
 $(-1)^{F}$ ( $F$ is the number operator for fermion) 
in the trace. 
Actually this modification is already mentioned in \cite{milo}.    
It interchanges  the untwisted and twisted sectors of 
the scalar fermion, and  the couplings become  
\eqbegin 
\chi^{c=-2}_{1}\chi^{a}_{(r+1/2)/q}, \hspace{3mm}\chi^{c=-2}_{\psi}\chi^{a}_{(r+1/2)/q},\hspace{3mm}
\chi^{c=-2}_{\sigma}\chi^{a}_{r/q},\hspace{3mm}\chi^{c=-2}_{\widetilde{\sigma}}
\chi^{a}_{r/q}.
\label{coup331}
\eqend 
These couplings are actually the same as the ones for 
the 331 state, whose internal 
degrees of freedom are given by the Dirac fermion. 
The partition function now equals to the one for the 331 state,$Z^{\rm ann}_{331}$.  
For the 331 state   
the modular invariance conditions are verified as in the Pfaffian
state and the fusion rules of quasiparticles 
are obtained correctly by the Verlinde formula.  
Actually the 331 state is a part of 
the generalized hierarchy \cite{wenzee} by the equivalence of 
the compactified boson at $R=1$ and the Dirac fermion.  
Thus the modular properties also follows from  \cite{cappelli2}. 

However,  the physical origin of insertion of $(-1)^{F}$ 
 or other maps of the scalar fermion to  the Dirac fermion is unclear. 
 Here we give another method which can be interpreted 
physically. We consider a half transformation of $(V)$ of (\ref{vcond}), 
\eqbegin
\zeta \rightarrow \zeta + \tau/2,
\eqend 
which we temporarily call $(X)$. 
Under the transformation $(X)$,  $\chi^{a}_{r/q}$ and 
$\chi^{a}_{(r+1/2)/q}$ transform as 
\eqbegin 
X : \chi^{a}_{r/q} \rightarrow e^{-\frac{\pi i}{2q}(2\R \zeta+\R
  \frac{\tau}{2})}\chi^{a}_{(r+1/2)/q}, \\ 
X : \chi^{a}_{(r+1/2)/q} \rightarrow e^{-\frac{\pi i}{2q}(2\R \zeta+\R
  \frac{\tau}{2})}\chi^{a}_{(r+1)/q}. 
\eqend  
Thus, under this transformation, the couplings in the Haldane-Rezayi 
state are transformed to (\ref{coup331}), the ones for the 331 state. 
For the partition function we thus establish the following 
identity: 
\eqbegin 
Z^{\rm ann}_{\rm HR}(\tau,\zeta+\frac{\tau}{2}) = Z^{\rm
  ann}_{331}(\tau,\zeta). 
\label{equivalence}
\eqend 
The physical interpretation of $(X)$ is the addition of 
a {\it half} unit flux quantum through the center of the annulus, 
which causes a spectral flow \footnote{The similar relation was also found 
in the calculation of the persistent edge current of the
Haldane-Rezayi state \cite{ino2}.}.     
This interpretation is transparent in the  modular transformed 
version of (\ref{equivalence}): 
\eqbegin
Z^{\rm ann}_{\rm HR}(\tau',\zeta'+\frac{1}{2}) = Z^{\rm
  ann}_{331}(\tau',\zeta'), 
\eqend 
where $\tau' =-\frac{1}{\tau}$ and $\zeta'=\frac{\zeta}{\tau}$.
This identity implies that the Haldane-Rezayi state 
 becomes a modular invariant consistent  state 
under the presence of a half unit flux quantum.   
In this picture, the Haldane-Rezayi state requires 
 such an additional flux for its  physical consistency. 
This observation  possibly gives  a clue  
for  ' The $\nu=\frac{5}{2}$ Enigma ' \cite{eisen}. 
Especially microscopic origin of the flux should be given.

Conversely, a 2+1 dimensional effective field theory
 of the Haldane-Rezayi state may be  realized as 
the  effective field theory of the 331 state with an additional flux i.e.  
two component Chern-Simons theory with a flux of 
a half unit quantum \cite{tobecontinued}.

For the Haldane-Rezayi state, as quasiparticles  involve  
a logarithmic operator \cite{gurarie},   
the fusion rules  in the bulk obviously cannot 
be obtained from  the modular matrix elements of the 331 state.  
The microscopic study \cite{read} supports 
that the topological orders in the Haldane-Rezayi state is 
indeed given by the $c=-2$ scalar fermion.      
Thus the computation of the topological order from 
the modular matrix elements  in Ref.\cite{cappelli2}
can not be extended to the Haldane-Rezayi state.

\section{Conclusions} 
In this paper, we first considered the spinon-holon FQH 
state constructed from a conformal block of the 
$SU(2)$ Wess-Zumino-Witten model at level $k$ 
and obtained the partition functions of the edge excitations 
on arbitrary multiply connected domains. 
We showed that  different sectorings for the holon 
edge excitations appear    
according to whether the level $k$ is even or odd. 
We then investigated the modular properties of the 
characters for nonabelian  FQH states.  We determined the 
modular properties of the U(1) characters which appear 
in nonabelian FQH states.  From them, we derived 
the modular properties of the characters for the spinon-holon state 
and paired states. We  confirmed 
 the extended modular invariance conditions 
for annulus partition functions. 
This enabled us 
to extend the relation between  the topological order and the modular 
matrix elements to nonabelian  cases.   
We also found that the extended characters for the Haldane-Rezayi state
 do not form a representation of the modular group.  We 
showed a new identity between the partition function for 
the Haldane-Rezayi state and the one for the 
331 state and suggested the  addition of half unit flux quantum 
 through
 the center of annulus 
to obtain the physically plausible partition function for 
the Haldane-Rezayi state.          
However,       
the relation between the modular properties and the topological order 
fails in the Haldane-Rezayi state 
even after the modification because of 
 the presence of a  logarithmic operator in the bulk.  
We also suggested that the appearance of the 
additional flux  may yield a clue for   'The $\nu=\frac{5}{2}$ Enigma'.

\vskip 0.2in
\noindent  
{\it Acknowledgement} \\ 
I would like to  thank  A.Cappelli, M.Flohr, D.Lidsky, 
J.Shiraishi and  G.R.Zemba for beneficial comments and encouragement.

\def\NP{{\it Nucl. Phys.\ }}
\def\PRL{{\it Phys. Rev. Lett.\ }}
\def\PL{{\it Phys. Lett.\ }}
\def\PR{{\it Phys. Rev.\ }}
\def\CMP{{\it Comm. Math. Phys.\ }}
\def\IJMP{{\it Int. J. Mod. Phys.\ }}
\def\MPL{{\it Mod. Phys. Lett.\ }}
\def\RMP{{\it Rev. Mod. Phys.\ }}
\def\AP{{\it Ann. Phys. (NY)\ }}

\end{document}